\documentclass[10pt,twocolumn]{article}

\usepackage{graphicx}
\usepackage{cite}
\usepackage{times}
\usepackage{algorithm}
\usepackage{algorithmic}
\usepackage{amsmath,amsfonts,amssymb,amsbsy}

\title{Circuits, Attractors and Reachability in Mixed-K Kauffman Networks}

\author{K.A.Hawick, H.A.James and C.J.Scogings \\
Institute of Information and Mathematical Sciences, Massey University, \\
North Shore 102-904, Auckland, New Zealand}

\date{15 November 2007}

\begin{document}

\maketitle

\begin{abstract}
The growth in number and nature of dynamical attractors in Kauffman NK network models are still not well understood
properties of these important random boolean networks.  Structural circuits in the underpinning graph give insights into
the number and length distribution of attractors in the NK model.  We use a fast direct circuit enumeration algorithm to
study the NK model and determine the growth behaviour of structural circuits.  This leads to an explanation and lower
bound on the growth properties and the number of attractor loops and a possible K-relationship for circuit number growth
with network size $N$.  We also introduce a mixed-K model that allows us to explore $N\left< K \right>$ between pairs of
integer $K$ values in Kauffman-like systems. We find that the circuits' behaviour is a useful metric in identifying phase
transitional behaviour around the critical connectivity in that model too.  We identify an intermediate phase transition
in circuit growth behaviour at $K = K_S \approx 1.5$, that is distinct from both the percolation transition at $K_P \equiv 1$
and the Kauffman transition at $K_C \equiv 2$.  We relate this transition to mutual node reachability within the giant
component of nodes.
\end{abstract}

\section{Introduction}
\label{sec:intro}

Kauffman's NK-Model~\cite{KauffmanNature69,Kauffman93} of an $N$-node random boolean network with $K$-inputs to a boolean
function residing on each node has found a significant role in the study of complex network properties.  Random Boolean
Network (RBN) models are effectively a generalisation of the 1-dimensional Cellular Automata model~\cite{WolframCA} and
have important applications in biological gene regulatory networks~\cite{KauffmanEtAlOnYeast} but also in more diverse
areas such as quantum gravity through their relationship with $\phi^3$-networks
~\cite{Baillie+JohnstonOnRBNGravity,Baillie+EtAl-Quantum-PhysicsLettersB}.  RBNs have many interesting properties
~\cite{Gershenson-Intro} and have been amenable to various analyses~\cite{Aldana+Coppersmith+Kadanoff} including
mean-field theory.  They also continue to be an important and interesting tool in studying biological and artificial life
problems~\cite{Fox+HillOnTopology,LynchOnRBN}.

\begin{figure}[htbp]
\begin{center}
\includegraphics[angle=0.0,width=8.5cm]{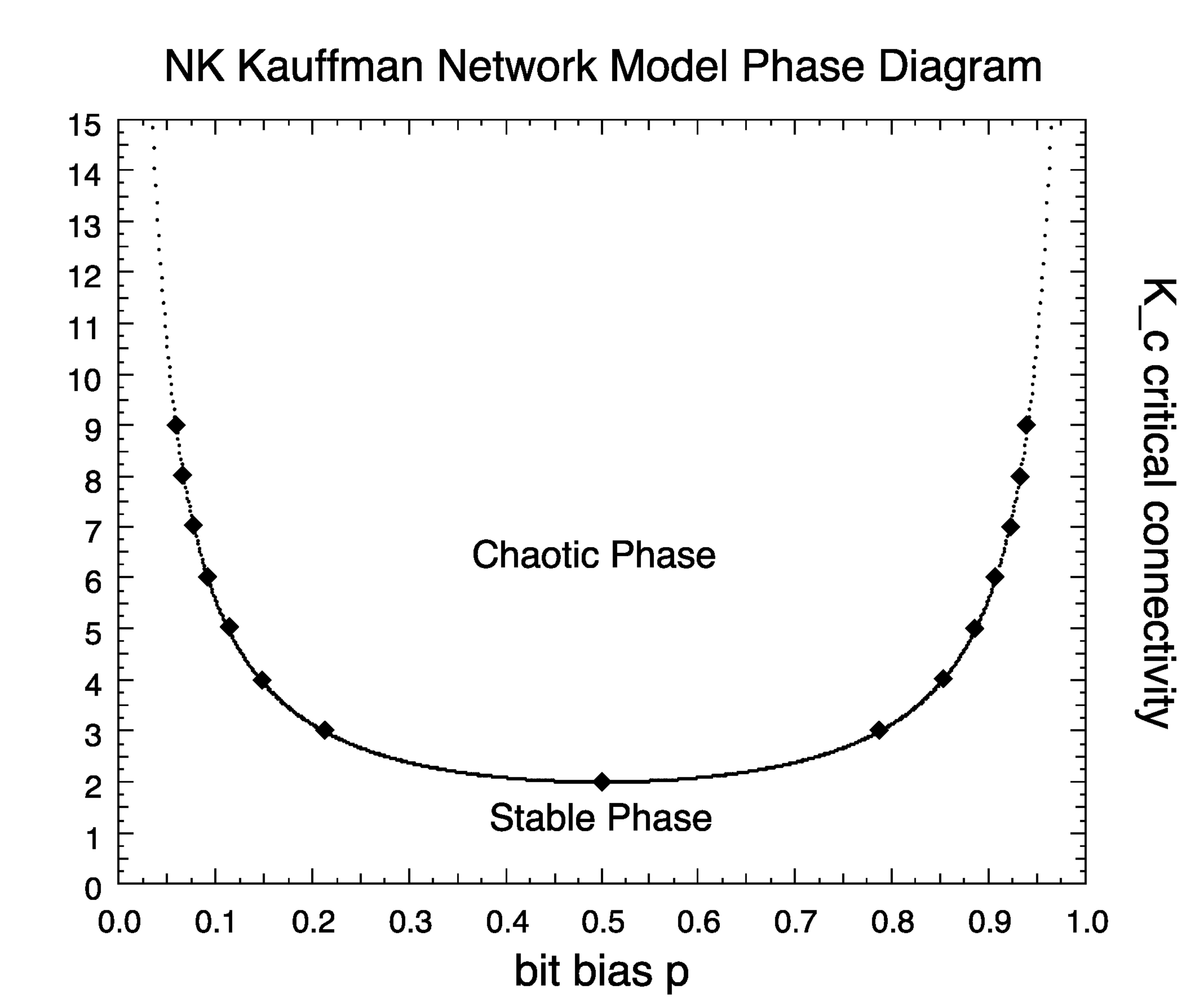} 
\caption{Phase diagram for the Kauffman NK Model in terms of (integer-valued) connectivity $K$ and bit-bias $p$. }
\label{fig:phase}
\end{center}
\end{figure}

One key property of RBNs is the now well established existence of a frozen phase and a chaotic phase
~\cite{Derrida+PomeauOnRandomNetworks,Derrida+StaufferOnKauffman} and the critical phase transition lies at the integer
value of connectivity $K_c = \frac{1}{2 p ( 1 - p)} = 2$ for unbiased networks with a mean boolean function output value
of $p=0.5$.  This gives rise to the phase diagram shown in figure~\ref{fig:phase} which is dominated by the
chaotic region and for which simulations are limited to the integer $K$ values shown.  It has therefore been of most
interest to study RBNs at or around this critical (integer) value of $K_c \equiv 2$.  In this paper we explore another
mechanism to explore the critical region by adopting a mixed-K system.

The Random Boolean Network or graph $G$ is expressed as a four-tuple $\mathbf{G} = (V, E, F, x)$ and has $N = |V| = |F|
= |x|$ nodes or vertices($V$), and $N_a = |E|$ directed edges or arcs, which express the $K_i$ inputs for node $i$.  Formally
we let $\mathbb{F}_{2} = \{0,1\}$ be a bit-field or Galois Field with the $\oplus$ operator defined modulo $2$ and
$\mathbb{F}_{2}^{K}$ is the set of all possible vectors of $K$ bits.  Vector $x \in {\mathbb{F}}_{2}^{K}$ and $x_i$ is then
the bit at position $i$.

The Kauffman NK-Model Network is constructed with fixed (integer) $K = 1,2,3,..$ and a boolean function $f_i \in F$ of $K_i$
inputs is assigned to each node. All the nodes of the network carry a boolean variable $x_i$ which may be initialised
randomly and each of which is updated (usually, but not necessarily) synchronously in time $t$ from its $j$-labelled inputs so that:
\begin{equation}
 x_i(t) \leftarrow f_{i}\left( x_{j}(t -1) \right), j=1,2,...,K_i
\end{equation}
The boolean functions $f$ thus map $\mathbb{F}_{2}^{K} \mapsto \mathbb{F}_2$. The mapping can be expressed as a truth
table and readily implemented as look-up table in a simulation program~\cite{CSTN-039}.

Work has been carried out on a number of different time-update mechanisms for boolean networks including asynchronous
algorithms~\cite{Harvey+Bossomaier-Time,Mesot+TeuscherOnRBN}.  In this paper we consider only synchronous updates where
all nodes execute their boolean function once, at the same time, at every time step.  Other studies have also considered
how noise~\cite{QuEtAlOnKauffman} effect the crossing times between distinct trajectories.  In this paper we
adhere to the quenched convention whereby node connections are assigned once and for all time, and particular boolean
functions are assigned randomly and uniformly to nodes once and for all time.

The NK-network model assigns the $K_i$ inputs for node $i$ randomly and with uniform probability across all possible nodes.  Even
for a large network there is still a non-zero probability of assigning a node as one of its own inputs. In the case of
$K_i > 1 $ there is also a possibility of assigning a node $j$ as an input of $i$ more than once. These self-edges and
multiple-edges can have a subtle effect on the behaviour of the NK-network model~\cite{HawickEtAlOnKauffmanCircuits}.

A significant body of work has now been carried out on the roles of different sub-classes of boolean functions including
the so called canalizing functions ~\cite{Szejka+DrosselOnCanalizing} and in particular the effect of bit-bias thresholds
and frozen or fixed-value boolean functions on particular elements of the network~\cite{Greil+DrosselOnThreshold}.  A
network can therefore be restricted to only have some subset of the possible boolean functions.  In the work we describe
here, we use network sizes large enough to sample all possible $K$-input boolean functions for the largest $K$ present
in the system.

An important consequence of the boolean functions in RBNs is the formation of attractor basins ~\cite{WuenscheOnAttractorBasins}.
These are observed in RBN models whereby diverse initial starting conditions will still lead to statistically similar
behaviour.  The state of the network falls into attractor cycles whereby a chain of interdependence of nodes (via their
boolean functions) leads to the network periodically repeating its state.  The number and length of these periods or
attractors is of great importance in understanding the behaviour of the NK-model and associated application problems.
This can be seen quantitatively by tracking a metric such as changes in the normalised Hamming distance between the
network's successive boolean states.  We discuss this metric in section~\ref{sec:pairmodel}.

\begin{figure}[htbp]  
\begin{center}
\includegraphics[angle=0.0,width=8.0cm]{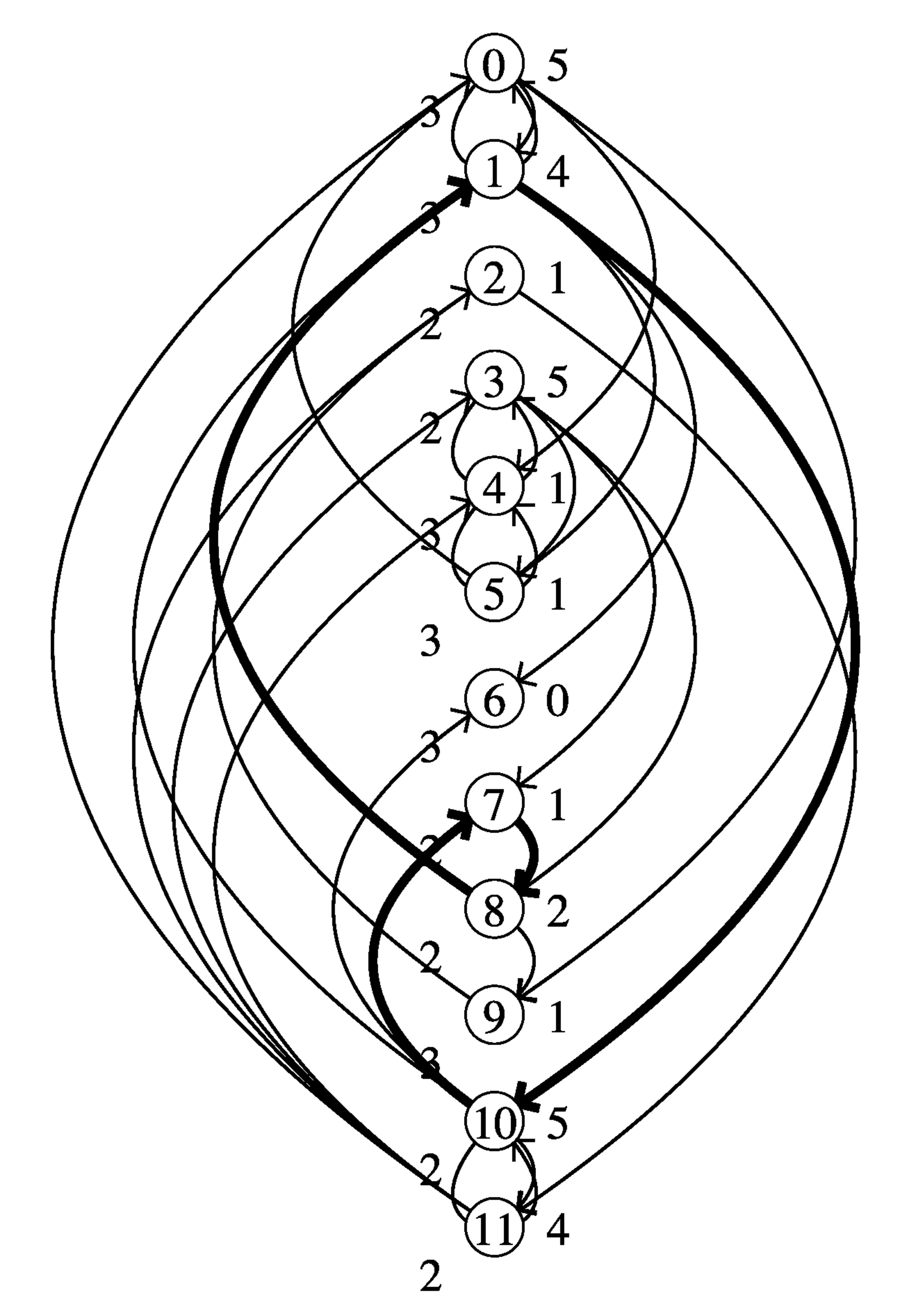}
\caption{$12$ Node Network, with $\left< K \right> \approx 2.5$.  For each node, input degree is shown on the left and
  output degree is shown on the right.  One particular circuit, 1-10-7-8-1, is shown. }
\label{fig:vchain}
\end{center}
\end{figure}

Of particular recent interest in the literature has been the uncertainty concerning the number of attractors
~\cite{DrosselOnAttractors,DrosselEtAlOnAttractors} and how their number and lengths varied with the size of the
network.  Scaling was initially believed to be $O(\sqrt N)$ ~\cite{Kauffman69}. It was later reported as linear
~\cite{Bilke+SjunnessonOnKauffman}, and then as ``faster than linear'' ~\cite{Socolar+KauffmanOnRBN} and subsequently as
``stretched exponential'' in ~\cite{Bastolla+ParisiOnRelevantElements,Bastolla+ParisiOnModularStructure} but is now known
to be faster than any power law ~\cite{Samuelsson+TroeinOnSuperpolynomial}.

A recent review of the RBN model ~\cite{Aldana+Coppersmith+Kadanoff} discusses the attractor behaviour in terms of the
loops of boolean variable states that form and several exact results concerning these loops have been obtained for the
case of connectivity $K \equiv 1$ ~\cite{Flyvbjerg+KjaerOnKauffman}.  Important observations concern the distribution of
components with particular sub-classes of possible boolean functions. These ``relevant elements'' are defined as those
nodes that are not frozen and that control at least one other relevant element in the system
~\cite{Bastolla+ParisiOnModularStructure}.  A number of important results have been obtained using particular sub-classes
of the possible boolean functions.  Drossel et al.~\cite{DrosselEtAlOnAttractors} have considered networks with non-fixed
boolean functions thus making all elements relevant and have therefore shown the equivalence of $K=1$ and $K=2$ networks
under appropriate restrictions on the boolean functions.

In this paper we use numerical methods to investigate the role that structural circuits play in the complex structure of
the network and the resulting attractor behaviour of RBNs both for mono-$K$ and mixed-$K$ network systems.  

Recent work in the literature has used trajectory sampling to study attractor behaviour.  The combinatorics of RBN
models means that the number of boolean functions grows as $2^{2^K}$ with a consequent rapid growth in the number of possible
network states with network size.  Taking limited numbers of sample trajectories through this state space can lead to
very misleading results.  Numerical sub-sampling of attractor trajectories seems to be the main difficulty behind
obtaining a good understanding of attractor scaling.  In this paper we explore the structural properties of RBNs
including the number and length distribution of elementary circuits and of components.  We compute these properties
exactly using brute force enumeration techniques for a range of network sizes and connectivities.  Our statistical
sampling is only over different randomly configured networks, not over attractor trajectories.

We have found it necessary to study quite large samples (up to 100,000) of networks of size up to 250,000 nodes.  The network
size $N$ must be at least large enough to adequately sample all possible $K$-input boolean functions for the largest $K$
present in our mixed systems.  There are some good software tools available for experimenting with RBNs such as those of
Gershenson~\cite{Gershenson-Intro} and Wuensche's DDLab~\cite{WuenscheThesis}, but we were forced to develop our own custom D
code to simulate very large-scale systems~\cite{CSTN-039}.

The so-called ``edge of chaos'' regime~\cite{Kauffman93} is quite narrow when $K$ is restricted to integer values.  In this
paper we also consider other ways to explore the model phase space.  We explore a mixed-$K$ system which we refer to as the
$N\left<K\right>$ Network model, with the understanding that although individual nodes (must) have an integer $K$ number
of inputs, the system as a whole can have a mean, or effective, $K$ value if there is a distribution of nodes each with
different number $K$ of inputs.  It then becomes a matter of deciding on a sensible K-distribution for a given model system.

Although some work has been done developing simulations that employ mixed-K models with Poisson or other distributions
~\cite{WuenscheThesis} we believe no one has yet studied mixed models in systematic detail and furthermore, that our
pair-wise model is a novel way of achieving a definite $K_{\mbox{eff}}$.  In ~\cite{SkarjaEtAlOnRBN} Skarja et
al. employed a skewed binomial distribution of $K$ values and found an enhanced tendency to orderliness (stability) but
based their study of attractors on trajectory sampling.

It is not altogether obvious what the meaning of a distribution in $K$ might mean if it has a long tail with
values at high and low $K$ relative to the mean, such as the Poisson distribution in node outputs that results from the
mono-K NK model.  Consequently we have chosen to study a pair-wise model that generates a $K$ value for each node that
interpolates between a pair of integer $K$ values.  We define a parameter $P$ that is the linear probability of a node
having $K=K_2$ rather than $K=K_1$.  So the cases $P=0$ or $P=1$ correspond to the pure integer states of all nodes having $K_2$ or
$K_1$ respectively, and $P=0.5$ constitutes an equal mix of the designated pair of $K$ values.  This means we can approach the
transitional value of $K=2$ from above or below and also means we can explore properties with K-dependent relationships
thoroughly over a larger range of fractional $K$ values rather than just the small subset of (small) integer $K$ values
that is feasible numerically.

Figure~\ref{fig:vchain} shows a small network generated with our mixed pair model.  Nodes have 2 or 3 inputs but can have
zero or more outputs depending upon the random distribution used during construction in the normal way for NK networks.

In~~\cite{HawickEtAlOnKauffmanCircuits} we explored the behaviour of the growth of the number of elementary circuits and
the resulting circuit length distributions in integer $K$-valued NK networks.  We suggested that the number of circuits
gave a lower bound on the number of attractors present in the associated NK network model.  Our numerical experiments
exactly enumerated the circuits in various NK networks, without resorting to trajectory sampling and we concluded that
the growth in the number of attractors had to be at least as fast as exponential in network size $N$.  This clarified
some of the recent controversy concerning the growth rates in the number of attractors.  

In our earlier work we had an insufficient range of $K$ values to determine a circuit growth relationship with $K$.  In
this present paper we have sampled higher $K$ values and also examined the pair-wise or fractional $K$ model over a large
range of $K$ values and have therefore been able to determine likely relationships between $K$ and the growth of the
number of circuits with network size $N$.

We have also observed long-time variations in the normalised Hamming distance for various K-valued networks at small and large
network sizes.  These suggest the very strong importance of adequately sampling all possible Boolean functions for a
given $K$ value.  This phenomenon may also explain the anomalous and misleading results on attractor growth obtained from
sampling trajectories in too-small networks.

In section~\ref{sec:pairmodel} we discuss the pair-wise $K$ model and implementation issues. We summarise the the role
circuits appear to play in Kauffman nets in section~\ref{sec:circuits} and their enumeration in
section~\ref{sec:enumeration}.  In section~\ref{sec:results} we present results on the number of circuits and their
length distribution both for integer and fractional $K$ network systems. In sections~\ref{sec:discussion} and
~\ref{sec:summary} we offer some discussion of the results and conclusions concerning circuit growth with $K$ and $N$ and
the properties of the mixed-pair $K$ model.

\section{$K$ Pairs and the $N\left<K\right>$ Model}
\label{sec:pairmodel}

The conventional Kauffman NK model with mono-K can be extended to a system with a distribution of K-input nodes.  RBN
simulation tools like Wuensche's {\tt DDLab}~\cite{WuenscheThesis} do make provision for this but it is unclear how to
systematically investigate possible distributions, particularly when large samples are required.  We might intuit that
the mean or effective $K$ value for the whole system plays an important role, but it is not clear how smeared-out any
behaviour might be that results from multiple $K$ values.  Our pair-wise model approaches the critical value in $K$ from
either side by adopting a simple uniform mix of just two possible values of $K$.  Most useful are $K=2,3$ and $K=1,2$ to
approach $K_c \equiv 2$ from above or below, respectively.

An effective-$K$ for the whole network can be defined as:
\begin{equation}
K_{\mbox{eff}} = \left<K_i\right>  = (1 - P ) . K_{\mbox{low}} + P . K_{\mbox{high}}  
\end{equation}
where any individual node $i$ has an integer valued number of exactly $K_i$ inputs and the average is over all $N$
nodes.  Individual nodes are randomly assigned (once and for all time) their particular $K$ value.

We can investigate both static and dynamic properties of this simple mixed model.  Static properties are measured from
simple graph analysis, and dynamics can be obtained by examining the Normalised Hamming distance between subsequent bit
states of the networks' nodes.

If we have the vector $x \in {\mathbb{F}}_{2}^{K}$ and $x_i$ is the bit at position $i$ in the network, we define the
Hamming weight as
\begin{equation}
w_H = | \{i|x_i \ne 0, i=1,2,...,K\}|
\end{equation}
and the Hamming distance between two vectors $x,y$ as 
\begin{equation}
H_d(x,y) = w_H( x \oplus y )
\end{equation}

It is useful to normalise this by the network size $N$.  Of particular interest is the Hamming
distance between subsequent states of the network.  This is easily calculated as $1-a_{\mbox{same}}$ where $a_{\mbox{same}}$
is the fraction of nodes that have the same bit value at subsequent steps or the single-step correlation function.

\begin{figure}[htbp]
\begin{center}
\includegraphics[angle=0.0,width=8.0cm]{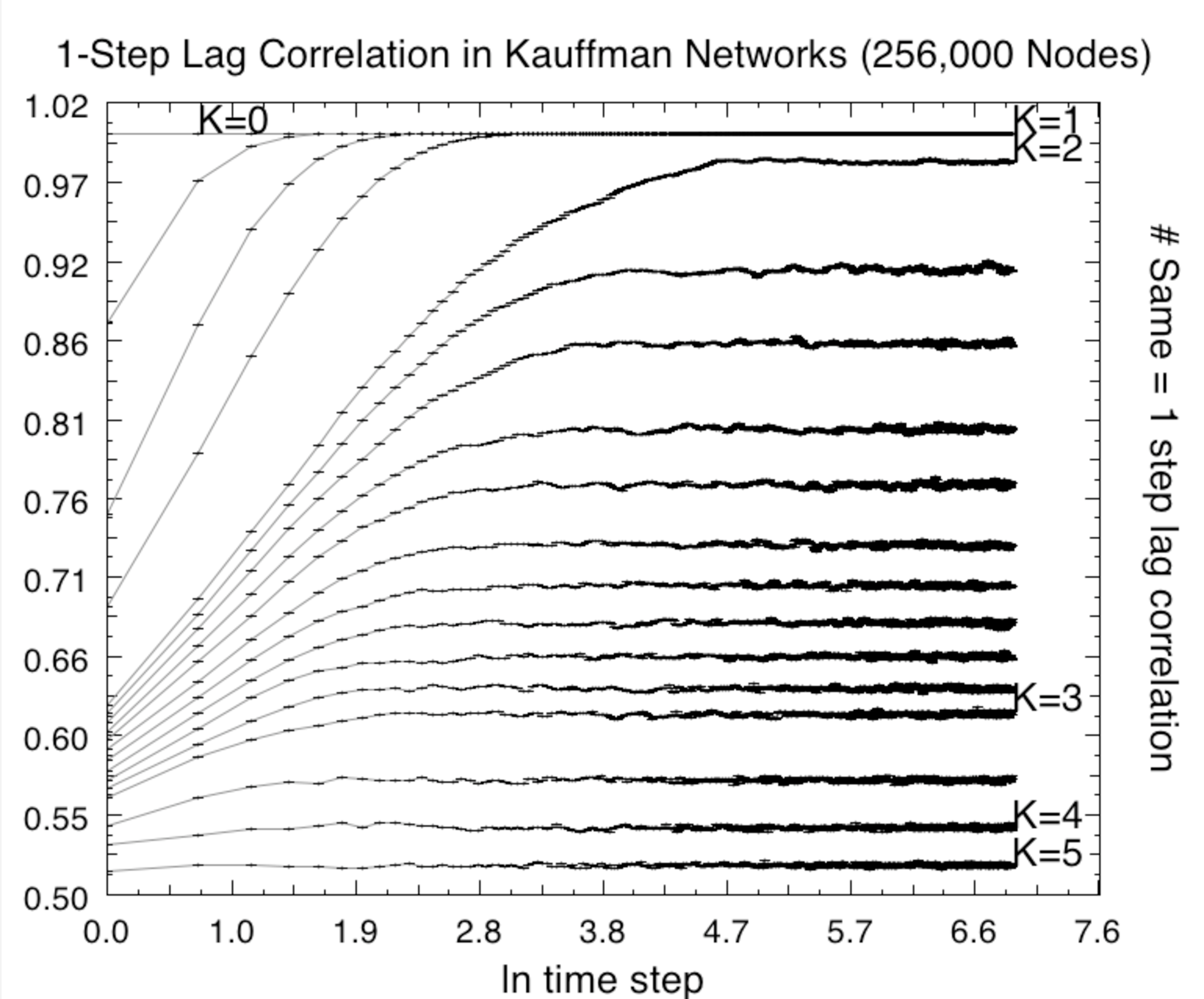} 
\caption{Single-Step lag correlation function (or fraction $a_{\mbox{same}}$) for $256,000$ Node Kauffman Networks for
  $K=0,1,2,3,4,5$ and networks with mixed pairs of $K$ values in between.  This is the same as ``one minus the
  normalised Hamming distance.''}
\label{fig:correlation}
\end{center}
\end{figure}

Figure~\ref{fig:correlation} shows the single step correlation function as measured from a variety of different K
values.  At low-K the network is barely connected and it very rapidly converges to a fixed Boolean state.  In
highly connected networks the attractor loops (of various lengths) introduce periodic cycles of correspondingly varied
frequencies.  Nevertheless the mean state of the network still converges to a stable value that depends critically upon $K$.

\begin{figure}[htbp]
\begin{center}
\includegraphics[angle=0.0,width=8.0cm]{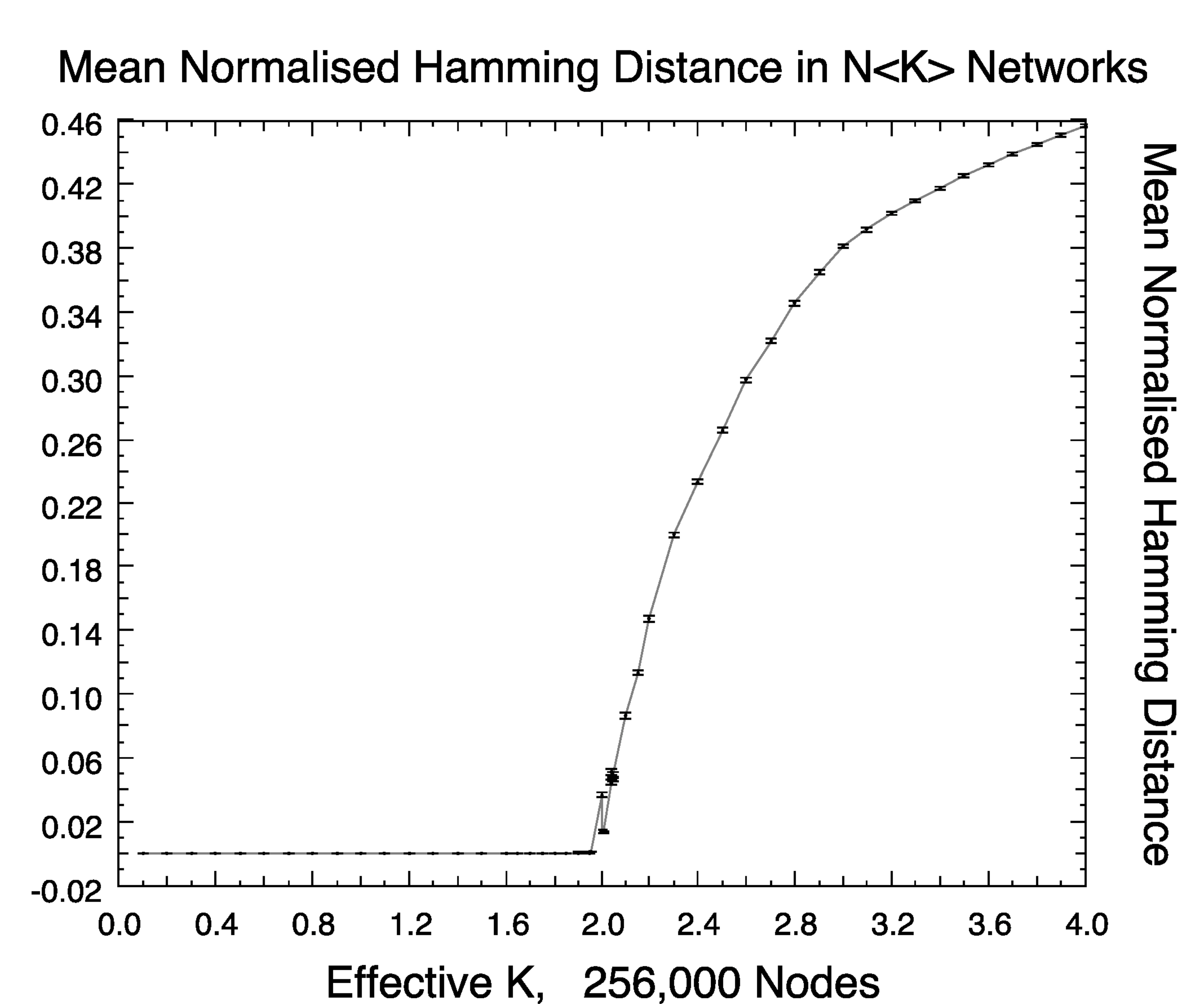} 
\caption{Normalised Hamming long term mean values}
\label{fig:mean-norm-hamming}
\end{center}
\end{figure}

Figure~\ref{fig:mean-norm-hamming} shows the long-term mean convergent values of the Normalised Hamming distance
$\lim_{t \rightarrow \infty} H_d^1$ between subsequent (ie time difference of 1) Boolean states of the network as it varies with
mean $K$ in our $N\left<K\right>$ model.  These are measured for 10 separate samples of an $N= 256,000$ node network.  We
observed that a stable reproducible long-term convergence value is obtained for each K-valued network only if $N$ is
significantly larger than the corresponding number $N_{B_f} = 2^{2^K}$ of possible Boolean functions.  Not unreasonably,
if $N$ is smaller than $N_{B_f}$, then the network instantiation is not adequately sampling the Boolean function space.
Empirically we found $N > 4 N_{B_f}$ suffices to ensure a reasonable sample.  Individual networks generally converge to
a stable value within a few hundred time steps.  To avoid possible lack of convergence we discarded the first $N$ time
steps and averaged over a subsequent $N$ steps.  The average has to be taken over a time larger than any likely attractor
traversal times present.

Figure~\ref{fig:mean-norm-hamming} does show a very clear transition at $K=2$ as expected for the integer-K $NK$ model, but also
indicates that the normalised Hamming distance changes smoothly for the $N\left<K\right>$ model as we vary $K$.  This
suggests our pair-wise model is indeed a useful way to investigate meaningful non-integer K-valued systems.  Intuitively
we might expect that $H_d^1( K-K_c )$ might have a straight-forward power or other closed function relationship with
$\left<K\right>$, but we were not able to obtain a good numerical fit to any obvious forms.

\section{Structural Circuits and Attractors}
\label{sec:circuits}

Figure~\ref{fig:vchain} shows a 12-node mixed-K network with $K_i \in {2,3}$, $K_{\mbox{eff}}=2.5$ and where the
construction algorithm has allowed self-arcs - in other words the inputs for each node have been chosen according to a
flat uniform distribution so they can connect to themselves.  The consequent self-edges allow self-inputs in the
corresponding RBN.  These are known to play a vital role in supporting the number of attractors.  A self-input or
``self-ancestor'' in the input dependence chain of boolean variables anchors the periodic or attractor
behaviour~\cite{Aldana+Coppersmith+Kadanoff} of RBNs.

We felt intuitively that the presence of structural circuits would also be vital to the periodic attractor behaviour and
as Drossel et al. have shown there are definite relationships between the number of attractors and the number of loops.
In fact, the number of structural circuits provides a lower bound on the number of possible attractors.  Consideration
of the exact number of enumerated circuits, and their distribution, gives insight into the controversy over the number
of attractors in RBNs.

An elementary circuit is a closed path along a subset of the edges of the graph such that no node, apart from the first
and last, appears twice.  The number of elementary circuits for a fully connected graph is bounded by 
\begin{equation}
\sum_{i=1}^{N-1} C_{N-i+1}^{N} (N - i)!
\end{equation}
as given by Harary~\cite{Harary+Palmer}.  This expression therefore represents the limit for the number of structural
circuits in an NK-network when $K \rightarrow N$.

Figure~\ref{fig:vchain} shows one such circuit or loop in the network structure.  In fact, exact enumeration (as shown
in figure~\ref{fig:enumerated}) indicates that there are 30 arcs and 28 circuits, the longest of which is of length 10.
It has 4 self-arcs (and hence two circuits are duplicated) and 3 multiple arcs.  The maximum number of outputs is 5 and
the minimum is zero. If self-edges are disallowed we would obtain a higher number of circuits present in the network.

\begin{figure}  
\begin{center}
{ 
\begin{verbatim}
  0  0
  0  1  5  0
  0  1 10  7  8  9  2 11  0
  0  1 10  7  8  9  2 11  3  5  0
  0  1 10  7  8  9  2 11  3  5  0
  0  1 10 11  0
  0  1 10 11  3  5  0
  0  1 10 11  3  5  0
  0  9  2 11  0
  0  9  2 11  3  5  0
  0  9  2 11  3  5  0
  0  9  2 11  3  7  8  1  5  0
  0  9  2 11  3  8  1  5  0
  0  9  2 11  0
  0  9  2 11  3  5  0
  0  9  2 11  3  5  0
  0  9  2 11  3  7  8  1  5  0
  0  9  2 11  3  8  1  5  0
  1 10  1
  1 10  7  8  1
  1 10 11  3  7  8  1
  1 10 11  3  8  1
  2 11  2
  2 11  3  7  8  9  2
  2 11  3  8  9  2
  3  3
  4  4
 10 10
\end{verbatim}
}
\caption{Enumerated circuits for the network shown in figure~\ref{fig:vchain}.}
\label{fig:enumerated}
\end{center}
\end{figure}

\section{Circuit Enumeration}
\label{sec:enumeration}

Various algorithms have been formulated to count the circuits in a graph but these either use infeasible amounts of memory or are
time exponential ~\cite{TiernanOnCircuits,TarjanOnCircuits} with a time bound of 
\begin{equation}
O( N. e(c+1) )
\end{equation}
We count circuits using a variation of Johnson's algorithm ~\cite{JohnsonOnCircuits} implemented in D.  For graphs of $N$ vertices, $e$
edges, $c$ circuits and $1$ fully connected component, Johnson's algorithm is time bounded in time by
\begin{equation}
O\left( (N + e)(c + 1) \right)
\end{equation}
and space bounded by $O(N + e)$. Unlike Johnson's algorithm our code copes with partially connected graphs without
resorting to the need to treat each of the possible $N_c > 1$ components separately~\cite{Hawick+JamesOnCircuits}. This is
still a highly expensive process since the number of circuits $c$ itself grows very rapidly with $(N,e)$.

In the graph literature the term loop is unfortunately sometimes used to describe a self-edge or a circuit of length 1.
In the NK-networks we study the number of self-edges is much less than $N$, even for low $K$.  However we do count them
and observe the effect of allowing them in the number of possible circuits and their length histograms.  We have
extended Johnson's published algorithm to cope with graphs with directed arcs (and not just bi-directional edges); with
multiple components; and self-arcs.  The computational complexity is not changed although we store some additional
book-keeping information to support arcs.  At worst this doubles the memory space required.  All the work we report in this
present paper is compute-time bound and not space bound.

On a modern (circa 2007) compute server with 4GBytes memory and a speed of 2.66GHz, we found it was entirely feasible to enumerate
circuits exactly in networks of up to $N \approx 140$ for $K=1,2$.  Smaller networks were required for higher $K$.  We
were able to count components quite easily up to networks of around $N \approx 20,000$.  We were able to exploit the
near-linear speed-up of parallel job-farming to average our exact enumeration/counting results over many independently
generated networks.

In a detailed investigation of elementary circuits in the graph structure of NK
networks~\cite{HawickEtAlOnKauffmanCircuits} we found numerical evidence for rapid growth of the number of circuits with
network size N, but were unable to determine a reliable numerical relationship for growth in terms of $K$.  This is
largely because with integer $K$ we are restricted to only a very few practical values.  It is only feasible to study
networks up to around $K \approx 9$.  This situation is not likely to change even with linear improvement in computer
speeds or other supercomputing techniques since the growth in the number of circuits with $K$ scales so rapidly.

Using the mixed model however, we are able to investigate intermediate values in K-space and attempt to find an
empirical relationship for circuit growth with $N$  for different mean values of $K$.

\section{Circuit Measurement Results}
\label{sec:results}

In this section we present results of various numerical experiments, exactly enumerating the circuits over independently
generated sample networks.  We emphasise that these data are not based on sampling attractor trajectories.

\begin{figure}[htbp]
\begin{center}
\includegraphics[angle=0.0,width=8.0cm]{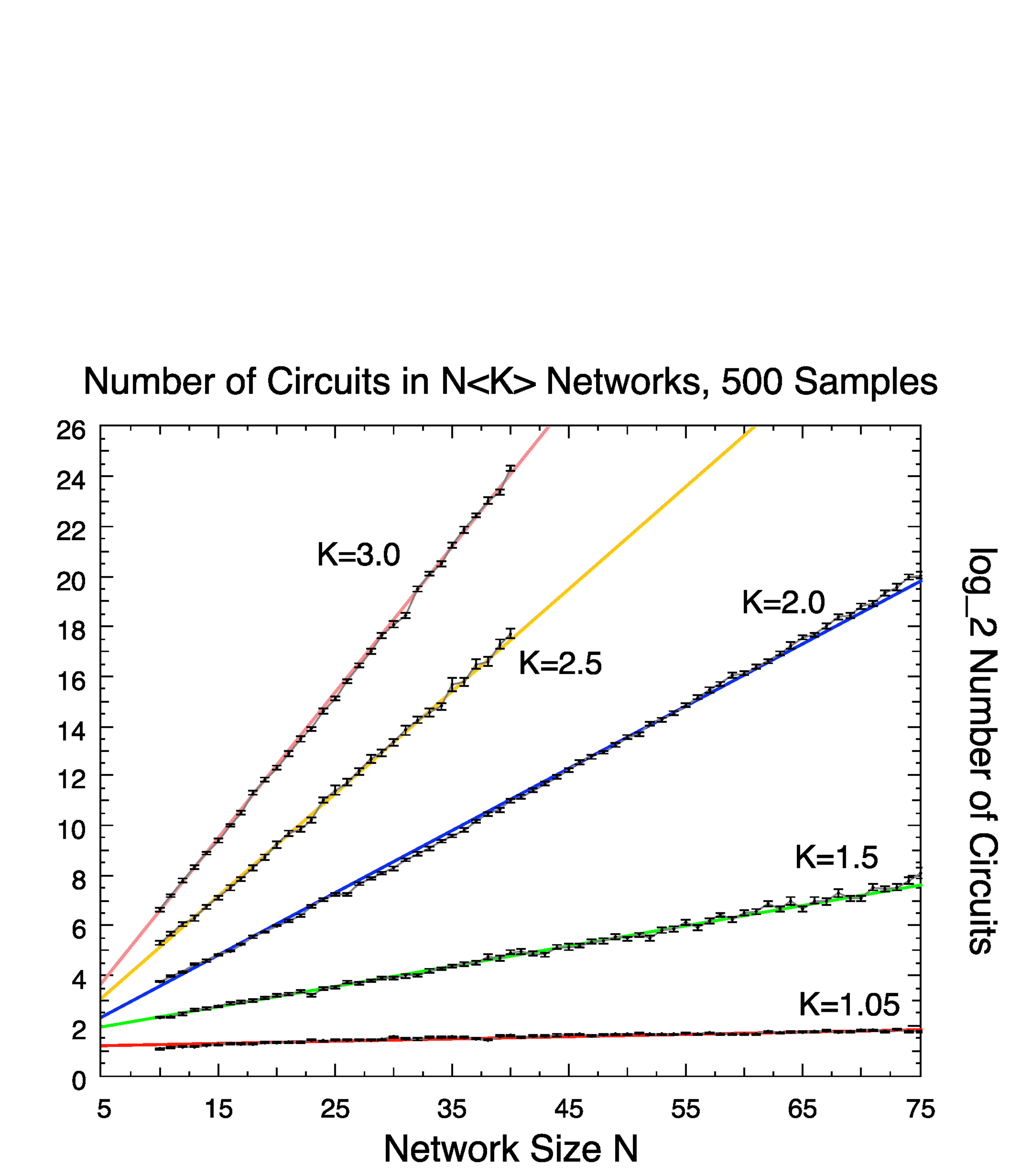} 
\caption{Number of circuits in Mixed-K Networks, sampled over 500 networks.}
\label{fig:ncircuits-A}
\end{center}
\end{figure}

\begin{figure}[htbp]
\begin{center}
\includegraphics[angle=0.0,width=8.0cm]{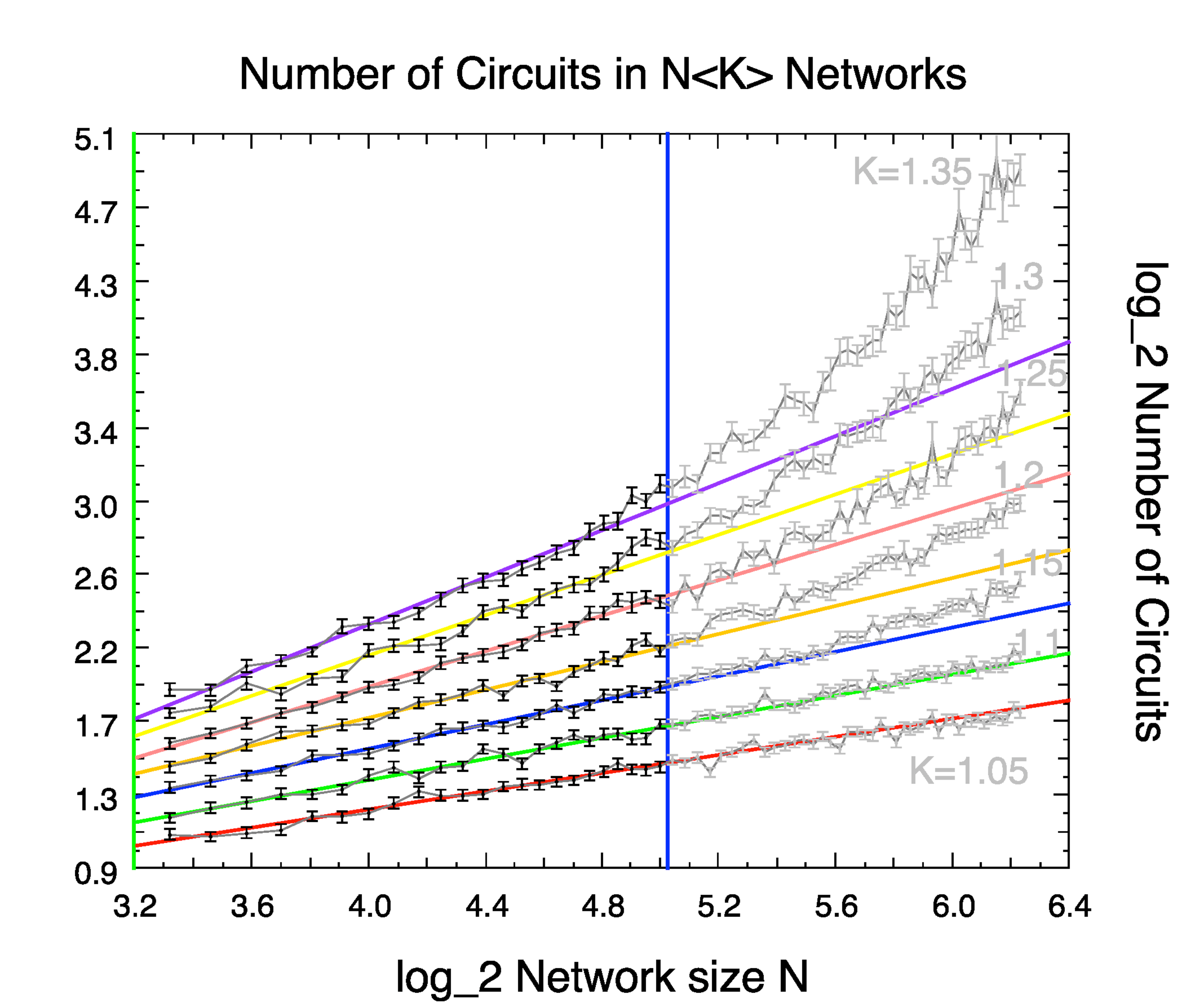} 
\caption{Number of circuits in Mixed-K Networks, sampled over 500 networks.}
\label{fig:ncircuits-B}
\end{center}
\end{figure}

Figure~\ref{fig:ncircuits-A} shows the number of circuits in mixed-K networks, sampled over 500 different networks.
Above a $K$ value of 1.5 a straight line fit to $\log N_c$ vs $N$ is a good model for the data, whereas for low K, there
appears to be a linear relationship between $\log N_c$ and $\log N$, as shown in figure~\ref{fig:ncircuits-B}.

As we conjectured in~\cite{HawickEtAlOnKauffmanCircuits} there is a definite change in behaviour between the low-K and
high-K regimes, however this transition is not a simple one arising from the percolation transition at $K=1$, as
figure~\ref{fig:ncircuits-B} indicates.

We can explore this effect further by examining the least-squares fitted slopes obtained from figure~\ref{fig:ncircuits-A}.

\begin{figure}[htbp]
\begin{center}
\includegraphics[angle=0.0,width=8.0cm]{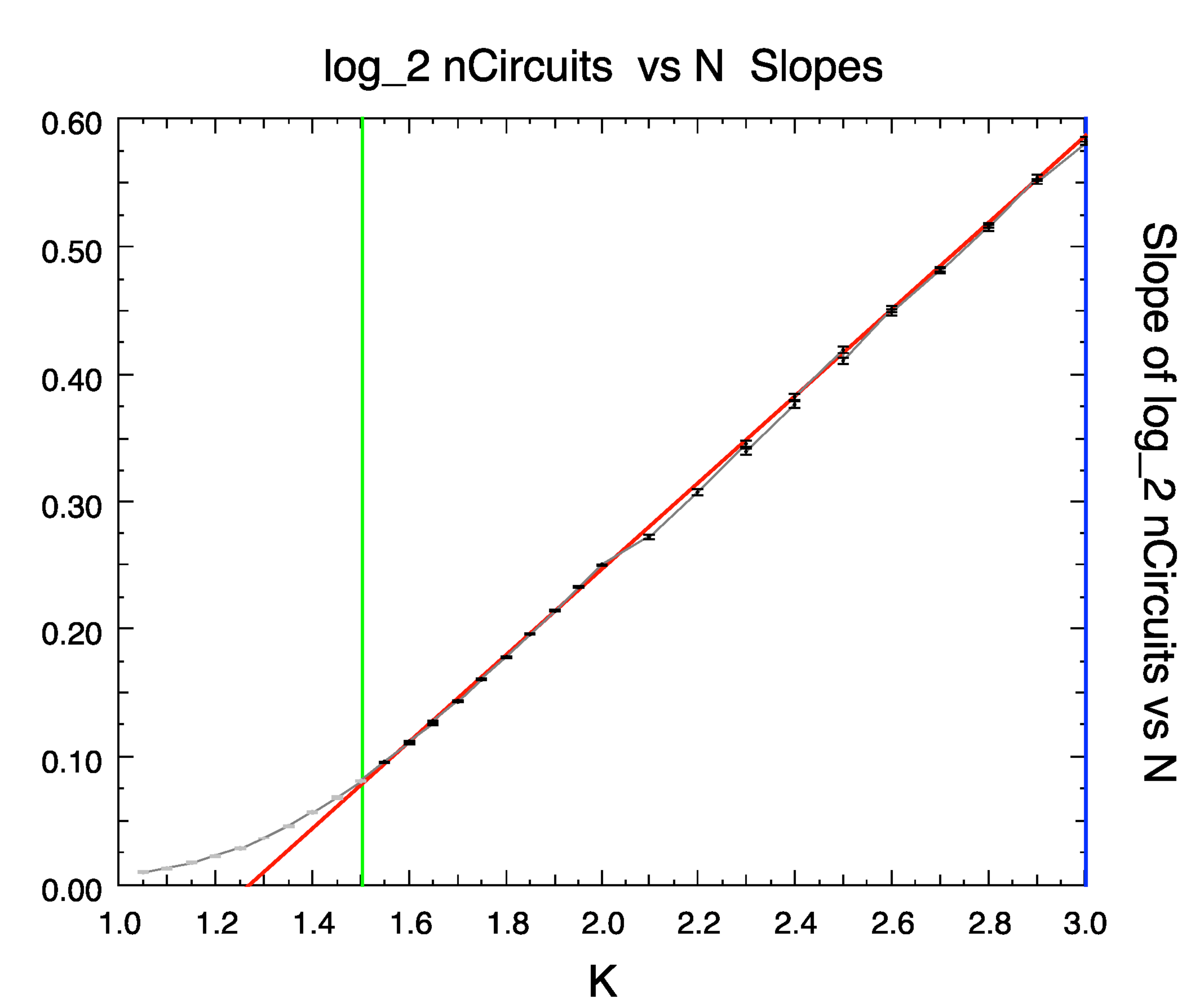} 
\caption{Circuit Slopes of $log_2 nCircuits$ vs $N$; fitted gradient is 0.340.  Experimental data follows the curve shown,
  with a dark fitted straight line.  Error bars were computed from standard deviations and are as shown.}
\label{fig:circuit-slopes}
\end{center}
\end{figure}

Figure~\ref{fig:circuit-slopes} shows a straight line fit to these fitted slopes from the data in
figure~\ref{fig:ncircuits-A} suggesting that the relationship $N_c \approx 2^{ s.N }$ holds where $s =s(K)$ and for
$K>1.5$ this is well fitted by a straight line so that $s \approx 0.340 K$ and hence $N_c \approx A_c 2^{0.340 K . N} $
holds.

Discounting the effect of self-arcs which grow in number linearly with $N$, and of multiple-arcs which grow with $N^2$,
the mean number of connections $N_A$ is approximately $K. N$ and therefore above $K=1.5$ the model data strongly supports
growth in the number of circuits with $N_A$

However, figure~\ref{fig:circuit-slopes} indicates a very clear departure from this behaviour at $K \approx 1.5$.
Interestingly the integer-K NK Kauffman model exhibits a transition in the long term Normalised Hamming distance at
$K=2$, as indeed does our mixed-K model as shown in figure~\ref{fig:mean-norm-hamming}.

As the error bars in figure~\ref{fig:circuit-slopes} show the data does support a straight line fit, although one could
convince oneself there is a small anomaly at the critical $K=K_c=2$.  This is not surprising given the pairwise nature
of our mixed model.  We are essentially approaching $K_c$ independently from above and below.  In the case of approach from above we
have a system whose nodes mostly have $K=2$ with a few of $K=3$ whereas from below we mix in a minority with $K=1$.  It
is in fact perhaps a point in favour of the simple pair-wise model that the two curves meet so closely.

It is also instructive to examine the distribution of circuit lengths present in a typical system.

\begin{figure}[htbp]
\begin{center}
\includegraphics[angle=0.0,width=8.0cm]{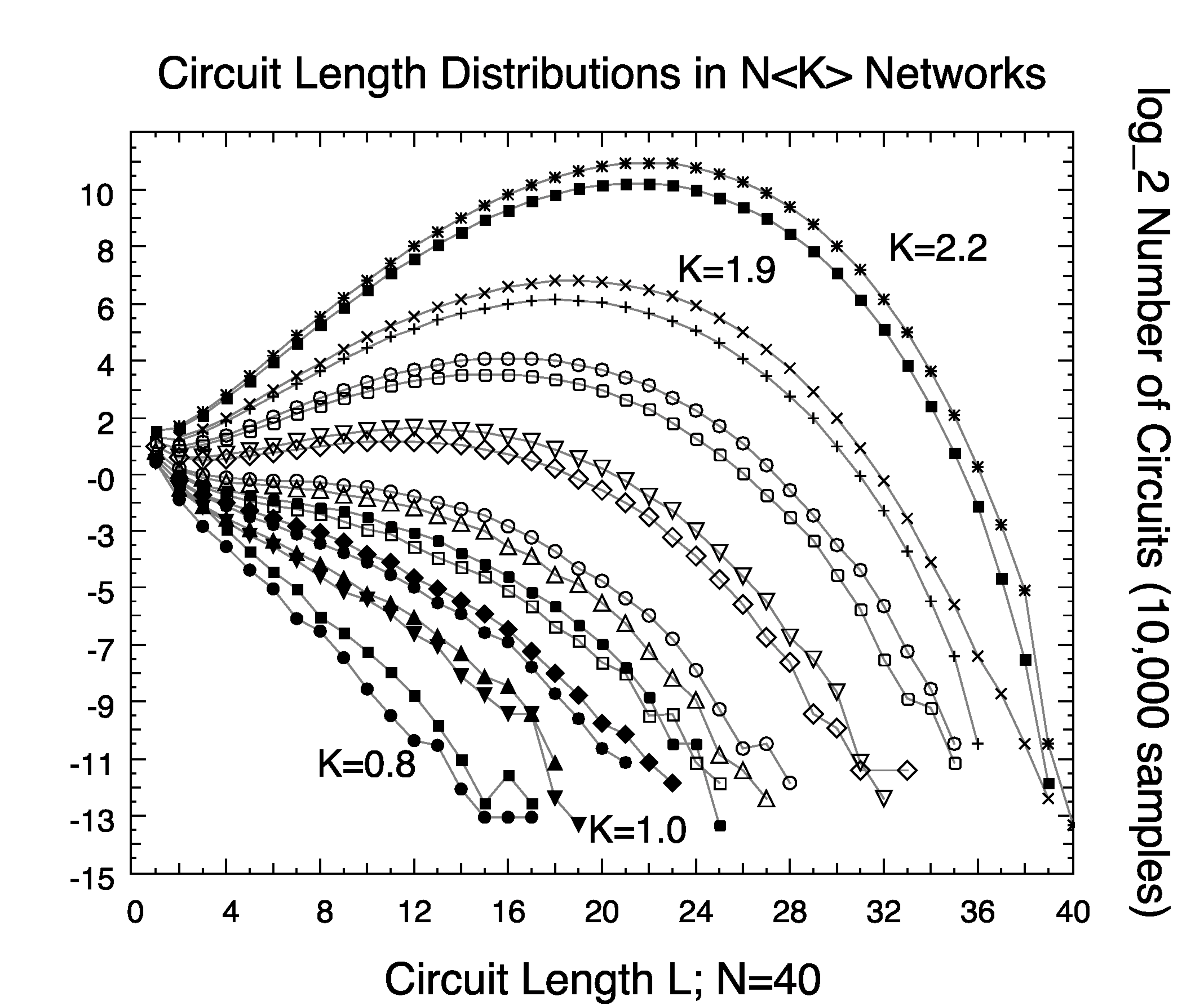} 
\caption{Distribution of circuit lengths in mixed $K$ systems, for fixed  network size $N=40$.}
\label{fig:frac-circuits}
\end{center}
\end{figure}

Figure~\ref{fig:frac-circuits} shows the length distribution of circuits in networks of fixed $N$ but different $K$.  At K
values above 1.5, the distribution of circuit lengths has a definite peak around $N/2$ and indicates that there is a non
zero, however small, possibility of Hamiltonian or near Hamiltonian circuits present in the system that include all $N$
nodes.  Below $K=1.5$ however the circuits length distribution falls monotonically with length and there is no modal
length.  Although we can only study relatively small network sizes in detail, we might expect that the fall off means
there would be almost no circuits of length greater than $N/2$ in large networks, and certainly, that the probability of
there being any Hamiltonian circuits is vanishingly small.

\begin{figure}[htbp]
\begin{center}
\includegraphics[angle=0.0,width=8.0cm]{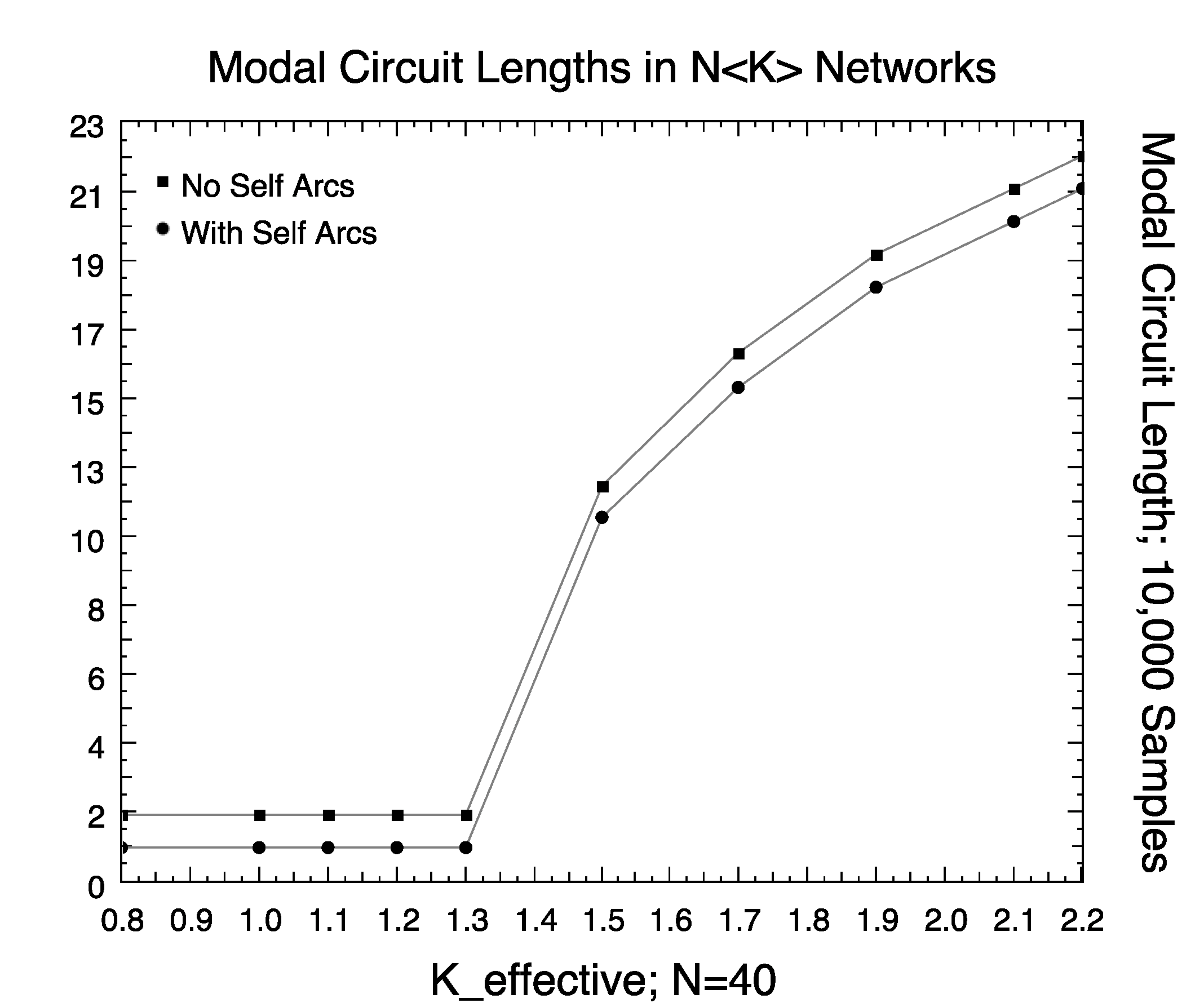} 
\caption{Modal Circuit lengths for 10,000 $N=40$ sample networks as it
  varies with $K_{\mbox{eff}}$.  Note separate curves for networks with and without self-arcs.}
\label{fig:lmax-keff}
\end{center}
\end{figure}

Figure~\ref{fig:lmax-keff} shows the modal circuit lengths as they vary with effective $K$ value.  Above $K=1.5$ the modal
circuit length is non-trivial and grows logarithmically with mean $K$.  Below the transition the most likely circuit
length present is unity for systems that allow self arcs, and two for systems that do not.

\section{Discussion on Network Composition}
\label{sec:discussion}

We have also carried out some standard graph metric analyses on our mixed system to clarify the role of the percolation
transition at $K=1$ on the growth circuits.  Theoretically we expect that, ignoring the effect of self-arcs and multiple
arcs, the percolation transition for infinite sized networks is exactly $K\equiv 1$
~\cite{CorrealeEtAlOnKauffmanPercolation}.

\begin{figure}[htbp]
\begin{center}
\includegraphics[angle=0.0,width=8.0cm]{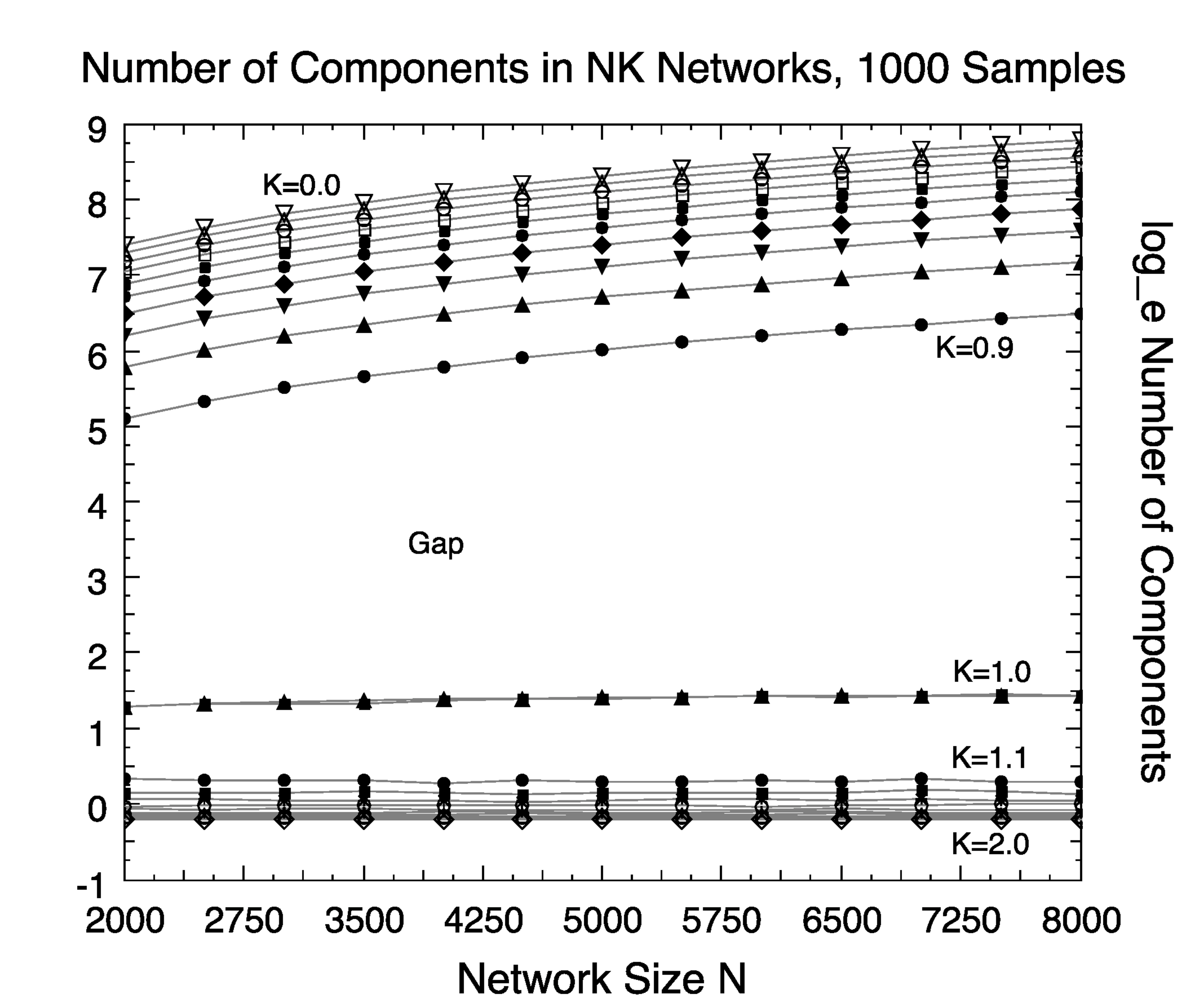} 
\caption{The number of cluster components in a mixed $K$ system for different $K$ and network sizes.}
\label{fig:log-nclusters}
\end{center}
\end{figure}

Figure~\ref{fig:log-nclusters} gives some insights into the composition of the system at a mixed value of $K$.  At high
$K>2$ the system is completely dominated by the single giant component and there are in fact practically no monomers.
For $N \gtrsim 2000, K \gtrsim 1$ the number of component clusters is almost identically unity and the number of
separate monomers almost identically zero.  This transition remains quite sharp in $K$ even for $N \gtrsim 100$ with a
clear gap below the critical $K=1$ value.

We found that the average number of monomers is still very small and does not vary with $N$ even for $K>1$.  A fully
disconnected system with $K \equiv 0$ has each node as a monomer and the number of monomers obviously then grows with $N$.
The first intermediate range of $0 < K < K_P$ shows the system become fully connected, and as discussed, $K_P
\rightarrow 1$ as $ N \rightarrow \inf $, but may be greater than unity for finite $N$.  The next intermediate range has $K_F
< K < K_c$ and shows some very interesting changes in the system's behaviour.

\begin{figure}[htbp]
\begin{center}
\includegraphics[angle=0.0,width=8.0cm]{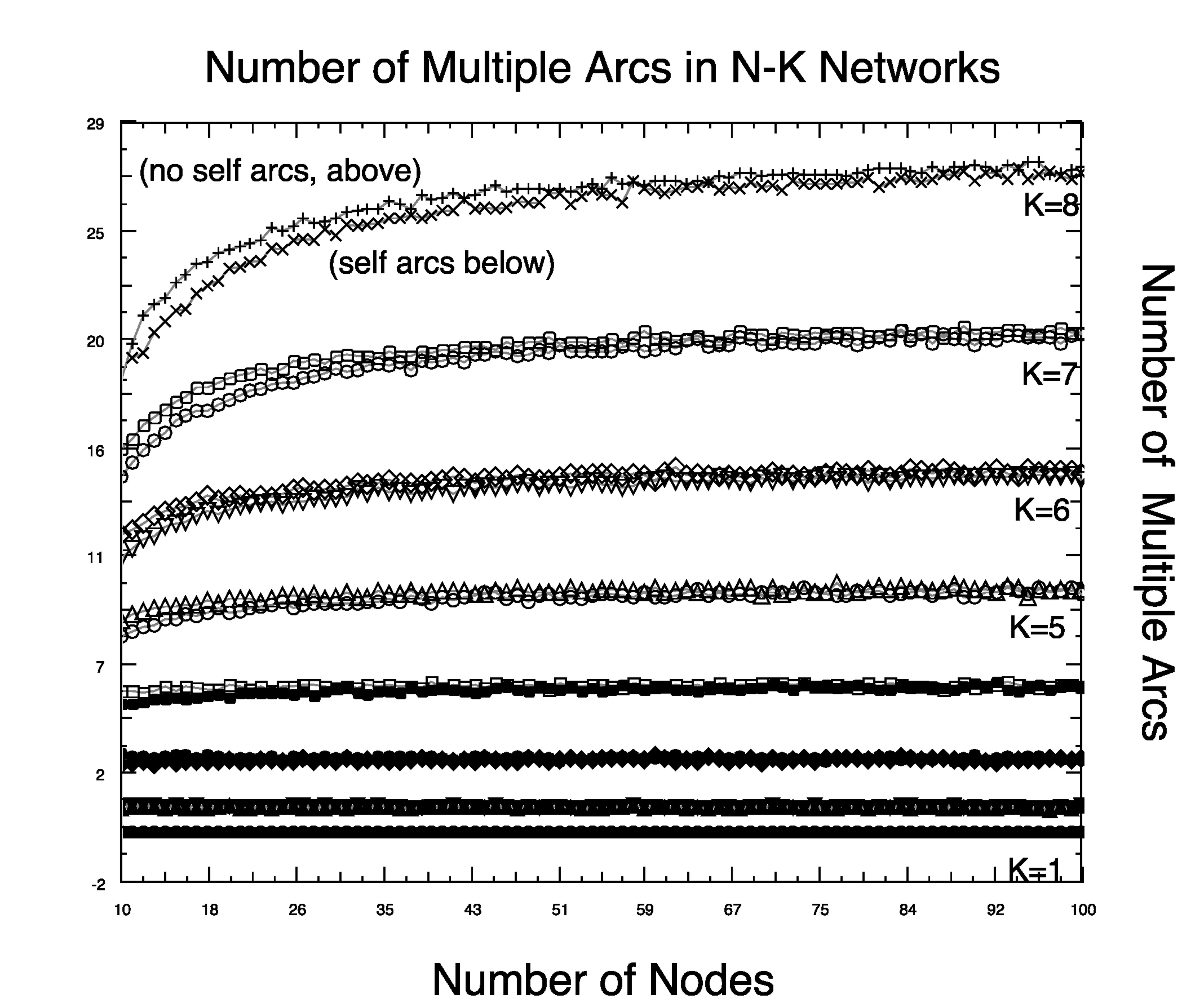} 
\caption{Number of multiple arcs in NK Networks vs network size $N$ for  $K=1,2,3,4,5,6,7,8$.}
\label{fig:nmultiple}
\end{center}
\end{figure}

We confirmed empirically our expectation that a system that is constructed to allow self-arcs will exhibit a growth in
their number linear in $N$.  In the work reported in this paper we have typically experimented with systems both with
and without self-arcs.  Figure~\ref{fig:nmultiple} shows a count of the number of multiple arcs and how these are in
fact influenced by the presence or absence of self-arcs.  For the low-K regimes we are most interested in, the number of
multiple arcs is almost invariant with network size, although this does grow logarithmically with $N$ -- particularly with
high-K.  It appears that neither self-arcs nor multiple arcs provide clues to the nature of the $K=1.5$ circuits
behaviour transition.

Another graph metric that is computationally inexpensive to compute is the all-pairs distance.  Elementary graph textbooks
illustrate this for fully connected systems and generally only for bi-directional graphs.

\begin{figure}[htbp]
\begin{center}
\includegraphics[angle=0.0,width=8.0cm]{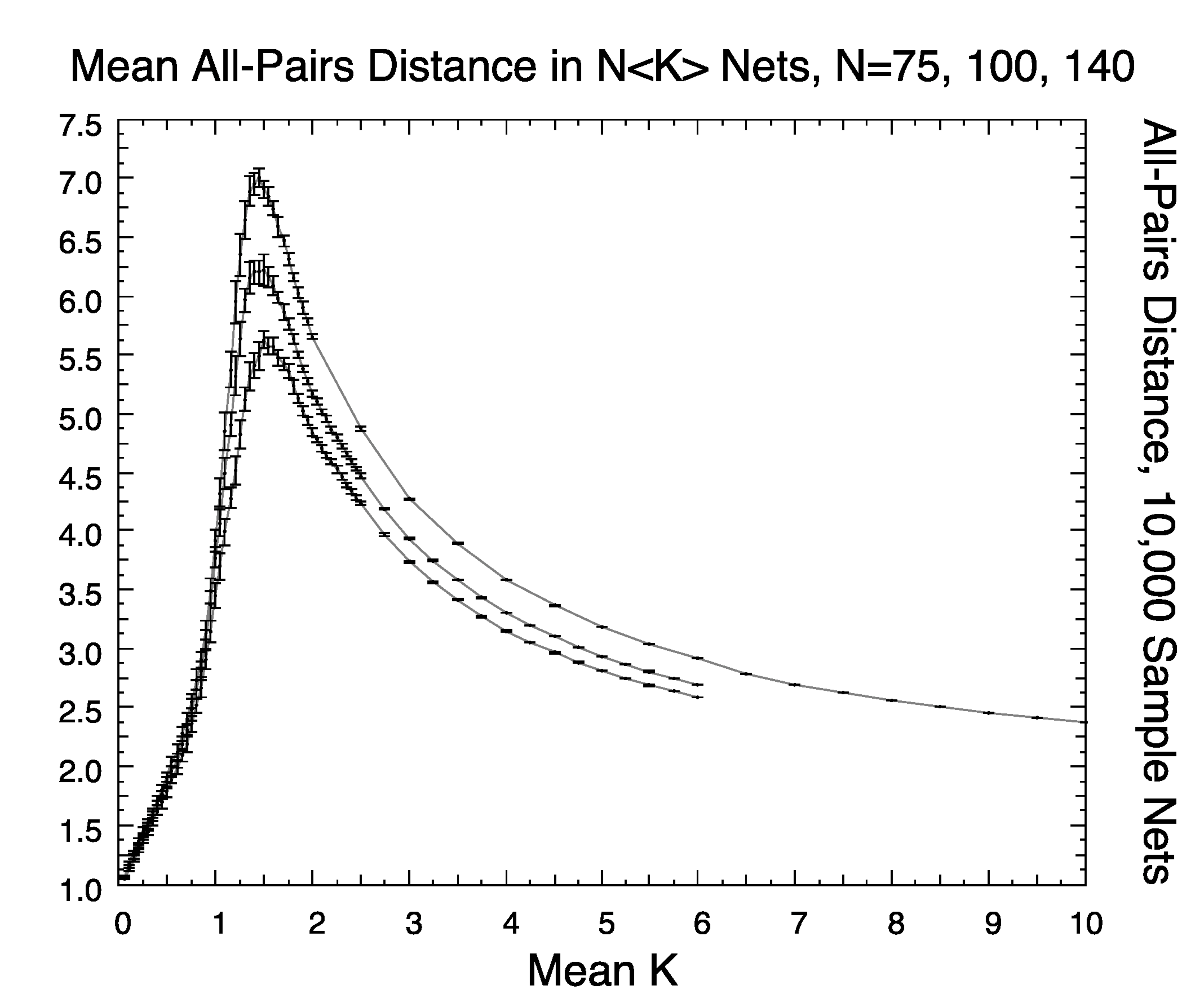} 
\caption{Variation of (modified) mean all pairs distance in mixed $K$ networks for $N=75, 100, 140$, showing maximal characteristic
distance  at $K=1.5$.}
\label{fig:all-pairs}
\end{center}
\end{figure}

Figure~\ref{fig:all-pairs} shows the significance of $K=1.5$ and needs to be carefully interpreted, recalling that it is
measuring a bulk graph property across regimes where the graph is neither necessarily fully connected nor fully mutually
reachable.  As $K$ rises from zero, the $N$ independent nodes become connected and the steepness of rise of the all-pairs
distance reaches a maximum at the percolation transition of $K=1$, however even when the network forms a single giant
cluster, the connections are directed and not every node is reachable from every other node.  Therefore a correct
weighted calculation of pair-pair distances, where we include two unreachable nodes as contributing zero rather than
infinity, highlights the structural role of connectivity $K=1.5$.  At this connectivity the network supports the greatest
set of traversal distances present consistent with being a fully connected system.  Note further that mutual
reachability tails off rather slowly and even at high $K$ there are still ``islands of directed disconnection'' in the
system.

We can further explore the notion of vertex reachability as distinct from connectivity.

\begin{figure}[htbp]
\begin{center}
\includegraphics[angle=0.0,width=8.0cm]{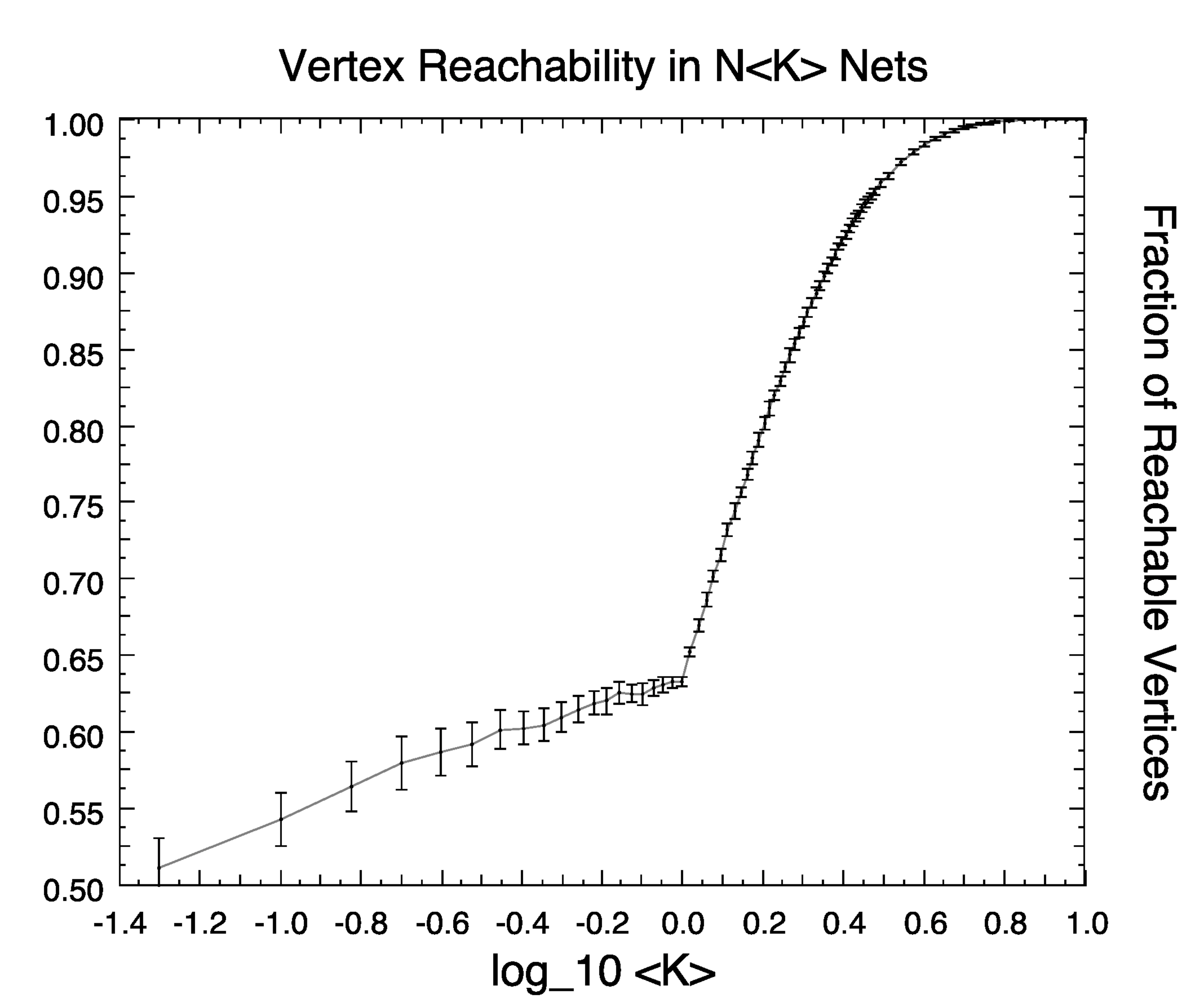} 
\caption{Fraction of Reachable vertices for different $<K>$ showing three regimes: multiple components for $K<1$; one component, but with varying vertex reachability  for $1 < K \lesssim 6$ and fully connect and fully reachable  for  large $K$. }
\label{fig:freachable}
\end{center}
\end{figure}

Figure~\ref{fig:freachable} shows the fraction of mutually reachable vertices for different mean $K$ in our model, with
560 nodes.  At $K<1$ we see multiple disconnected components, but we do not in fact reach full reachability until $K
\approx 6$.

Figure~\ref{fig:all-pairs} also shows the sharpening of the transition with increasing network size.  It is worth noting
that many of the applications of Kauffman networks as they relate to real physical and biological systems have very
definitely finite network sizes $N$ of a few hundred to a few tens of thousands and we are not solely interested in
thermodynamically sized systems. 

We can verify the bluntening of the percolation transition by counting the fraction of the network nodes in the
giant component.

\begin{figure}[htbp]
\begin{center}
\includegraphics[angle=0.0,width=8.0cm]{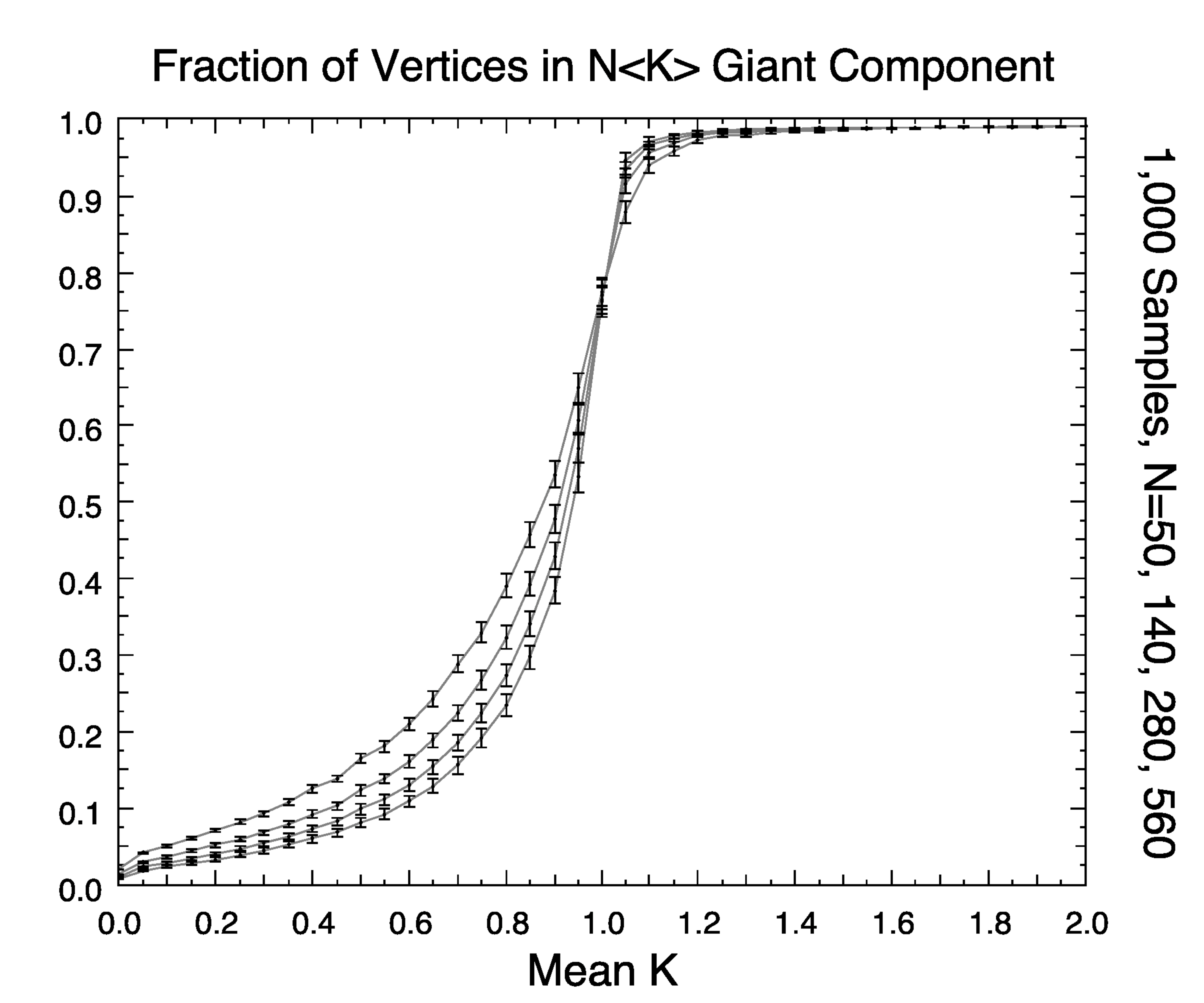} 
\caption{Fraction of vertices in the giant component for different $<K>$ showing a sharpening of the edge for increasing network size $N$.}
\label{fig:fgiant}
\end{center}
\end{figure}

Figure~\ref{fig:fgiant} shows how our network system generation algorithm will produce disconnected clusters even above
$K=K_P\equiv 1$ for finite network size $N$.  We conjectured that the $K=1.5$ transition might be related to partial
component disconnection, but even when one edits out everything except the giant components and arranges them by size, we
still see this effect within a fully connected component with mean $K =1.5$.

\section{Summary and Conclusions}
\label{sec:summary}

There seems to be several important transition values in the mixed-K model: the percolation transition at $K \approx
1$ from a disconnected to a fully connected system; the structural transition at $K  = K_S \approx 1.5$ and the Kauffman
transition from the stable to chaotic regime at $K_C \equiv 2$.

Our study was originally motivated on the assumption that there is a close relationship between the number of elementary
circuits in the underpinning structural network and the number of attractors that can be supported in an associated RBN.
As Aldana and Cluzel observe~\cite{Aldana+CluzelOnRobustNetworks}, the average connectivity $K_{\mbox{eff}}$ appears
irrelevant in describing highly heterogeneous scale-free topological regimes of an NK system, but it does appear to be
useful in characterising the transitional regime between $K_P$ and $K_c$.

Specifically, it appears that the structural transition at $K=1.5$ strongly influences the number of possible circuits
in the system, but it requires the $K=2$ connectivity transition for the boolean functions to be able to exploit it and
to produce a number of attractors that cross into the chaotic phase.

In ~\cite{DrosselEtAlOnAttractors} Drossel et al. speculate that the vast number of attractors in $NK$ models appear to be
a consequence of the synchronous updating scheme.  While no doubt the synchronicity plays a role, we believe our work
shows that the growth of attractors is also a more fundamental consequence of the structural circuits present in the
underpinning $NK$ graph.

We have shown that above $K_S$ the number of structural circuits appears to be a relatively simple exponential function
of the number of connections $N_A$ where discounting self-arcs and multiple arcs, $N_A \approx K . N$.  This
relationship models both the NK Network system and our pair-wise $N\left<K\right>$ system.  The number of circuits
$N_C$ therefore does appear to be a useful lower bound on the number of attractors in both models, and it remains for
future work to refine this relationship and bound.

\section*{Acknowledgments}
We thank U.Scogings for invaluable assistance in proof-reading this document and the Allan Wilson Centre for use of the
Helix cluster supercomputer.
\bibliographystyle{unsrt}
\bibliography{046-arxiv}
\end{document}